# Extremely metal-poor gas at a redshift of 7

Robert A. Simcoe[1], Peter W. Sullivan[1], Kathy L. Cooksey[1], Melodie M. Kao[1,2], Michael S. Matejek[1], and Adam J. Burgasser[1,3]

**In typical astrophysical environments, the abundance of heavy elements ranges from 0.001-2 times the solar concentration. Lower abundances have been seen in select stars in the Milky Way's halo[1–3] and in two quasar absorption systems at redshift $z$=3 (ref 4). These are widely interpreted as relics from the early universe, when all gas possessed a primordial chemistry. Before now there have been no direct abundance measurements from the first Gyr after the Big Bang, when the earliest stars began synthesizing elements. Here we report observations of hydrogen and heavy element absorption in a quasar spectrum at $z$=7.04, when the universe was just 772 Myr old (5.6% its present age). We detect a large column of neutral hydrogen but no corresponding heavy elements, limiting the chemical abundance to < 1/10,000 the solar level if the gas is in a gravitationally bound proto-galaxy, or <1/1,000 solar if it is diffuse and unbound. If the absorption is truly intergalactic[5,6], it would imply that the universe was neither ionized by starlight nor chemically enriched in this neighborhood at z~7. If it is gravitationally bound, the inferred abundance is too low to promote efficient cooling[7,8], and the system would be a viable site to form the predicted but as-yet unobserved massive population III stars in the early universe.**

We observed the recently discovered $z$=7.085 quasar ULAS J1120+0641 (ref 6) in January 2012 with the FIRE infrared spectrometer[9] on the Magellan/Baade telescope. Our data provide a 12-fold increase in spectral resolution over the discovery spectrum at similar signal-to-noise ratio, enabling study of weak heavy element absorption lines that are diluted by the instrumental profile at lower resolution.

Our spectrum (Figure 1) confirms the presence of unusually strong Lyman alpha (Lyα) resonance absorption from neutral hydrogen (H I) in the immediate foreground of the quasar. This absorption is clearly visible as a contrast between the observed flux (black) and an intrinsic source spectrum model (red) at λ < 0.98 μm. However, the data also fall well below the source template at wavelengths redder than the Lyα transition at the quasar's systemic redshift (λ > 0.9829 μm). This has been interpreted[5,6] as a Lorentzian damping wing of the H I Lyα line at redshift very close to that of the quasar, indicating a high neutral hydrogen column density[10]. Such absorption could be caused by a long column of low density, intergalactic hydrogen with high neutral fraction in the vicinity of the quasar. Or, it could arise in compact, high density gas gravitationally bound to an early galaxy. Such proximate damped Lyman alpha (DLA) absorption systems have numerous analogs at lower redshift.

At z = 5 and below, every known absorption system with sufficient neutral hydrogen to elicit damping wings also exhibits absorption from heavy element lines[11–13]. However we find no evidence of heavy element absorption despite the sensitivity of the FIRE data, which is sufficient to detect metals (elements heavier than helium) at abundance levels characteristic of lower redshift DLAs. We do detect narrow metal absorption lines from highly ionized gas at the redshift of the quasar, manifested in C IV and N V. However these are offset from the damped H I absorption by Δv = +711 km/s (equivalent to 800 proper kpc if Δv is purely cosmological), and there is substantial flux transmission at the associated H I wavelength for these lines. The heavy element lines are therefore most likely internal to the quasar host itself and not physically coincident with the neutral gas.

[1] MIT-Kavli Institute for Astrophysics and Space Research. [2] Department of Astronomy, California Institute of Technology.
[3] Center for Astrophysics and Space Science, University of California, San Diego

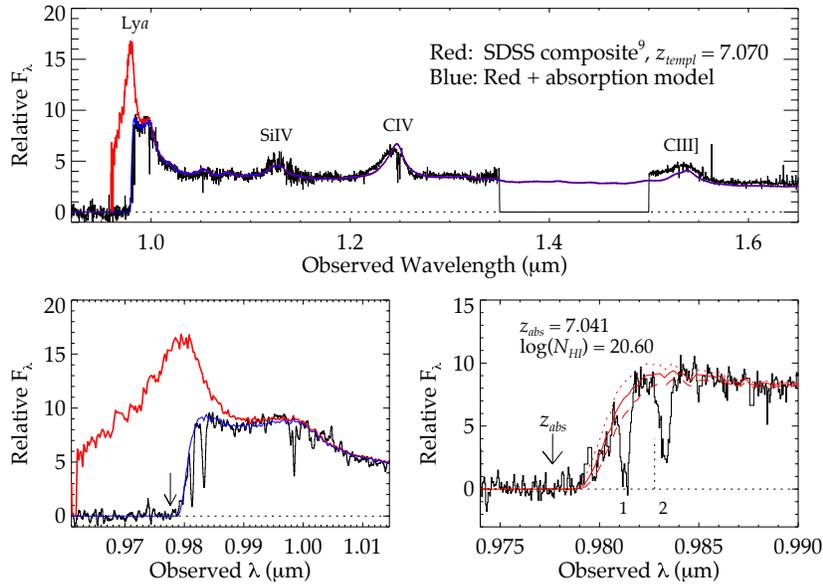

**Figure 1**: FIRE spectrum of ULAS J1120+0641, alongside our estimate of the intrinsic source spectrum and a composite model including foreground absorption. The unabsorbed continuum us shown in red, and the blue curve includes the absorption. The continuum is constructed from a composite of quasars in the Sloan Digital Sky Survey[14,16]. C IV absorption intrinsic to the quasar host galaxy is seen to the red of the labeled CIV emission peak. However, the C IV emission line is anomalously blueshifted[6] in ULAS J1120, so we compute the redshift distance between the absorber and quasar host using its MgII[6] or [C II] (ref 25) redshift. *Bottom Left:* The Lyman alpha region of the spectrum with unabsorbed continuum model (red) and absorbed continuum (blue). The vertical arrow marks the location of Lyα absorption at $z = 7.04$. *Bottom right*: Detail of the damping wing with HI absorption fit. The quasar's emission redshift[25] (7.0842) is indicated with the vertical dashed line. Two additional optically thin Lyα absorbers (labeled 1 and 2) are apparent in the quasar's near zone at $z = 7.0721 \pm 0.0001$ and $z = 7.0855 \pm 0.0001$ ($\Delta v = -424, +161$ km / s from the host, see Supplementary Information). These data have not been continuum normalized, so a slight downward slope is visible toward redder wavelengths.

Quantitative chemical abundance estimates are usually impossible for *z* > 5.5 quasar absorbers because the benchmark neutral hydrogen line is severely blended and saturated in the forest of neighboring Lyα systems. However the damping wing near the emission redshift of ULAS J1120 offers a unique opportunity to measure its H I column density. In conjunction with upper limits on the heavy element column density, this yields a straightforward upper limit on the chemical abundance of metals.

The H I column density estimate is sensitive to the detailed shape of the damping profile, which is fitted to the ratio of emitted to observed flux (the ratio of the red to black lines in Figure 1). This ratio depends critically on how the intrinsic (i.e. unabsorbed) shape of the quasar's Lyα emission line is modeled, including both its absolute flux density and its redshift, which fixes the location of the emission peak. The details of this procedure are described in the Supplementary Online Material, but to summarize, we experimented with several different prescriptions, including four different quasar composite spectra generated from low redshift surveys[14–17], and additionally a principal-component analysis fit[18] extrapolated over the Lyman alpha region. For each of these continua, we calculated the H I column density required to produce the damping wing redward of the systemic Lyα line via Voigt profile model fitting, finding a best-fit value of log($N_{HI}$)=20.60 cm$^{-2}$. For any one continuum model, the formal fit error for log($N_{HI}$) was of order 0.02-0.03 dex, but the true error is much more likely to be dominated by systematic uncertainty in the continuum. By experimenting with different choices of continuum and absorber redshift, we estimated the range of allowable $N_{HI}$ as 20.45-21.0, at a best-fit absorption redshift of z=7.041$^{+0.003}_{-0.008}$ (95% confidence).

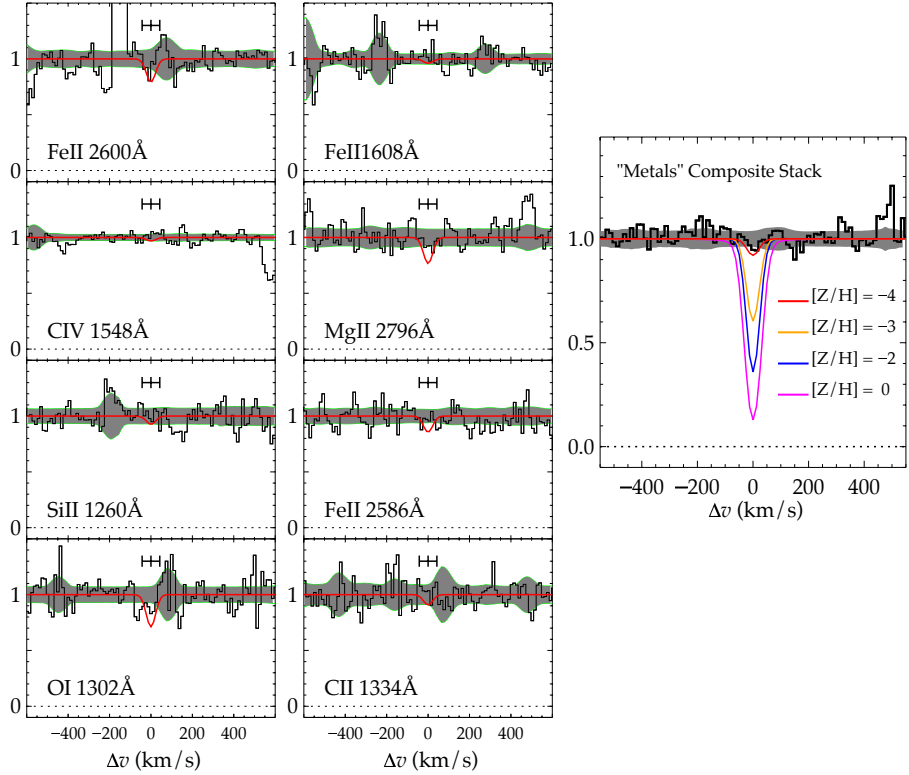

**Figure 2:** Continuum-normalized spectral regions where key heavy element transitions would fall at z=7.041 if the absorber were typical of a low redshift DLA. 1σ error contours are shaded grey, and the extraction aperture is indicated with horizontal bars. A Voigt profile with $b$ = 10 km/s and $N$ set by our column density upper limit is shown in red for each panel. Velocity offsets are relative to the rest frame of the HI absorber. *Right Panel:* Composite stack of all heavy element transitions, generated using an inverse-variance weighted mean and solar relative abundances. Each transition is scaled to the cross section and relative abundance of O I. Overlaid curves show the predicted profiles for varying metallicity levels. The stack shows no statistically significant absorption, though there is a fluctuation at the 1σ level, corresponding to an effective [O/H] < -4.

We estimated upper limits to the metal line column densities both by curve-of-growth analysis and by direct Voigt profile fitting (Table 1, see also online Supplemental Information). For systems with log($N_{HI}$) > 20.3, the transitions in Table 1 represent the predominant ionization states for their respective elements[19,20]. The one exception to this is C IV, which is secondary to C II but we include as an ionization constraint (discussed below). We therefore follow the usual practice for DLA systems and do not apply ionization corrections when estimating abundances[19,20].

Our strongest abundance limit is derived from Si II, which yields a 1σ (2σ) upper bound of 1/20,000 (1/10,000) the solar metallicity. With such unusual abundances one must consider the possible effects of ionization, particularly given the proximity of the brightest known object in the z = 7 universe. However, prior studies of other proximate damped Ly$\alpha$ systems indicate that ionization plays only a minor role and may be counterbalanced by a tendency for such systems to have higher than average metallicities[13,21]. Additionally, our spectrum places strong constraints on the lack of ionized gas seen in C IV, which would yield a ten-fold lower abundance limit than C II if it dominated the ionization balance (on account of its higher signal-to-noise ratio and larger atomic cross section). This rules out an ionized phase as a major source of missing metals in this absorber.

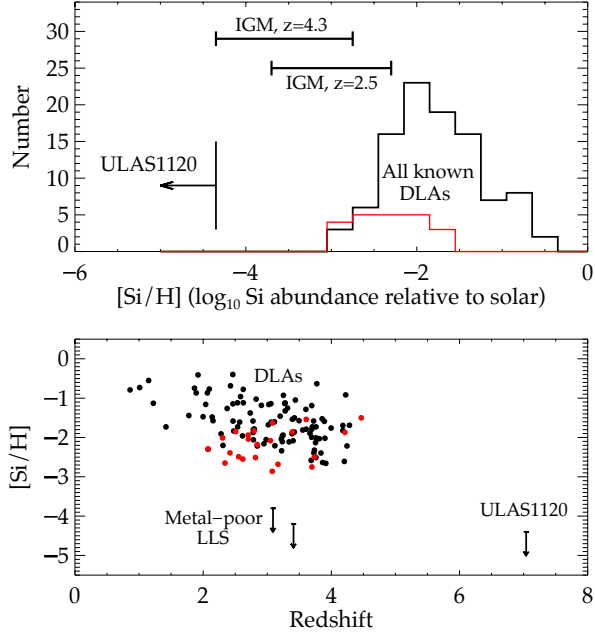

**Figure 3**: Comparison of z = 7.04 abundance measurement in ULAS J1120 with existing values from the literature. The top panel displays a histogram of the Silicon abundance for known DLAs[11,12,21], while the bottom panel shows the same systems at their respective redshifts. The top panel includes the +/-1$\sigma$ abundance ranges for intergalactic absorption systems[22,23] at redshift 2.5 and 4.3. The bottom panel also shows the locations of two recently reported metal-poor Lyman limit QSO absorbers[4] at z ~ 3. DLAs at lower redshift exhibit -2.5 < [Si/H] < -0.5,[11] with the lowest previously known system having [Si/H] = -2.75.[13] Surveys designed expressly to uncover metal-poor DLAs (shown with red histogram and points) find -2.86 < [Si/H] < -1.5.[12]

The limits presented above assume that the absorbing gas resides in a compact structure that is well-represented by a discrete line or cloud, which is appropriate for DLAs at lower redshift. However at $z = 7$ the absorption could also result from the integrated contribution of diffuse intergalactic gas if the neighborhood of ULAS J1120 has not yet been ionized by starlight. Using Monte Carlo simulations of intergalactic H I and heavy element absorption (described in the Supplementary Material), we find that for physical conditions leading to an intergalactic damping wing our spectrum still restricts heavy element absorption to the < $10^{-3}$ solar level, even with no DLA.

These chemical abundance limits have significant implications for either of the two physical scenarios considered. If the diffuse absorption model is correct, then at $z = 7$ the intergalactic medium must be both metal-poor and substantially neutral even in the neighborhood of a bright quasar. Such intergalactic material would not yet have mixed with the polluted interstellar byproducts of galaxies, so it should not be surprising for its heavy element content to be small. At later epochs (redshifts $z = 2$ to 4) heavy elements are actually observed in intergalactic space, with abundances distributed log-normally between 1/300 and 1/3,000 solar[22,23]. Our z = 7 limit excludes the upper half of this distribution but enrichment at the low end could elude detection in the FIRE data. Nevertheless, if this one object proves to be representative of the universe at these epochs, is plausible that redshift seven predates both the radiative and the chemical feedback thought to be hallmarks of reionization.

Alternatively, the observed hydrogen may sample the interstellar medium of a very early proto-galaxy. In this case, our limit of < $10^{-4}$ solar abundance is an order of magnitude lower than any previously measured in a predominantly neutral gas reservoir (Figure 3). Comparably low heavy element abundances have recently been reported in two Lyman limit systems at $z \sim 3$ (ref 4), but the gas in those environments is at most 0.8% neutral[4], and their total neutral hydrogen column is 2-3 orders of magnitude smaller. In contrast, the system producing the ULAS J1120 damping wing is at least 10% neutral and possibly much more so[5]. Figure 3 shows that metal-poor gas clouds represent a small percentage of all known systems at lower redshift[11,12,22]. At $z = 7$, the first chemical abundance measured in any non-AGN environment would already be as pristine as the most metal-poor absorption systems at $z = 3$ and below.

DLA gas is commonly said to be a fuel supply for star formation, but our limits on the abundance in this geometry are sufficiently low that normal gas cooling channels for Population I and II star formation would be suppressed. Theoretical calculations suggest that below an abundance of [C/H] ~ -3.5, the normal fine-structure atomic cooling mechanisms lose their effectiveness, and the predominant mode of star formation occurs through molecular $H_2$ cooling, resulting in high-mass Population III stars[7,8]. Taken together with its high redshift, this fact renders the $z = 7.04$ absorber in ULAS J1120 a viable site for population III star formation if the neutral gas is organized into a bound halo.

We caution that many more observations of $z > 7$ absorption systems will be required to establish whether the trends of large neutral hydrogen column and low metal content are representative of the overall universe at this early epoch. Indeed the existence of heavy elements in both emission and absorption in the background quasar host indicate that local enrichment is underway in some environments. However the confluence of neutral gas and extremely low chemical abundance, taken together with the young age of the universe (772 Myr) suggests that current observations may already be reaching the era corresponding to the onset of star formation and cosmic chemical enrichment.

**Acknowledgements:** We thank John O'Meara and Anna Frebel for helpful comments during the preparation of this paper. M. Haehnelt also provided very useful advice on methods for modeling the quasar near-zone, and G. Richards shared his composite QSO spectra in electronic form. This work includes data gathered with the 6.5 meter Magellan Telescopes located at Las Campanas Observatory, Chile. R.S. acknowledges support from the NSF under awards AST-0908920 and AST-1109115.

**Author Contributions**: R.S. constructed the FIRE instrument; and together with P.S. designed and executed the observations, performed the analysis, and prepared the manuscript. K.C. prepared observations and edited the manuscript. M.M. assisted with the pipeline software used to reduce the spectroscopic data, and M.K. wrote the software to perform eigenspectrum continuum fits. A.B. contributed to the spectrograph construction, and executed observations for the program. All authors helped with the scientific interpretations and commented on the manuscript.

**Competing Interests Statement**: The authors declare that they have no competing financial interests.

**Correspondence**: Correspondence and requests for materials should be addressed to R.Simcoe. (simcoe@space.mit.edu)

| Ion | Abundance |
|---|---|
| O I 1302 Å | [O / H] ≤ -3.39 (-3.25) |
| C II 1334 Å | [C / H] ≤ -4.16 (-3.86) |
| Si II 1260 Å | [Si / H] ≤ -4.35 (-4.03) |
| Fe II 2586 Å | [Fe / H] ≤ -3.09 (-2.92) |
| Fe II 2600 Å | [Fe / H] ≤ -3.43 (-3.41) |
| Mg II 2796 Å | [Mg / H] ≤ -3.89 (-3.74) |
| Stacked Composite | [Z / H] ≤ -4.07 (-3.90) |

**Table 1:** Direct abundance upper limits for five individual elements together with the limit for the stacked composite profile, under the critical assumption that the absorbing gas is organized into a gravitationally bound halo. Quoted values are 1σ upper limits, with 2σ limits shown in parentheses. Values reflect the logarithm of the elemental abundance (by number) relative to solar, calculated using $[X/H] = \log(N_x) - \log(N_{HI}) - \log(X/H)_\odot$ and standard solar abundance values[24]. If the absorption instead arises in unbound intergalactic gas, we derive upper limits of [X/H] < -3.0 using simulated sensitivity calculations described in the text.

# SUPPLEMENTARY INFORMATION

Extremely metal-poor gas at a redshift of $z \sim 7$

## S.1 Data

We observed ULAS J1120+0641 (hereafter ULAS J1120) over six different nights with the FIRE infrared echelle spectrometer[9] on the Magellan/Baade telescope, obtaining 37 separate exposures for a total integration of 15.01 hours. All data were taken with a 0.6" slit oriented at the parallactic angle for measured spectral resolution of R=6000 or 50 km/s. During this time the seeing ranged from 0.4 to 1.1 arcseconds FWHM so slit losses were variable. Between the individual exposures, which typically lasted 25 minutes but ranged from 15-30 minutes as scheduling required, we nodded the object along the slit to sample the detector and split cosmic rays. We interspersed observations of A0V standard stars at 60-90 minute intervals throughout the observations for telluric and flux calibration.

Each exposure was extracted individually and corrected for terrestrial atmospheric absorption using a custom-developed IDL pipeline. The software produces a 2-D estimate of the sky spectrum using pixels away from the object on the slit for background subtraction purposes[26]. A separate 1-D trace is extracted for each exposure using optimal spatial weighting, and this spectrum is corrected individually for telluric absorption and flux calibrated using routines from the Spextool software package[27]. Finally, we generated a master stack of the full data set with each exposure weighted by its squared signal-to-noise ratio. This optimally-weighted composite spectrum exhibits signal-to-noise ranging from 12 to 20 per 12.5 km s$^{-1}$ pixel in the regions of interest.

## S.2 Continuum Estimation

The damping wing near $z = z_{em}$ in ULAS J1120 is unique even among high redshift quasars; fundamentally this is what enables us to estimate $N_{HI}$ and hence measure chemical abundances. But the existence of this deep absorption also complicates estimation of the quasar continuum. At lower redshift the standard practice is to define the continuum through interpolation between minimally absorbed peaks in the spectrum on the blue wing of the Lyman alpha profile and throughout the Lyman alpha forest. In ULAS J1120, fully saturated Gunn-Peterson absorption eliminates these high-transmission peaks. The continuum must therefore be extrapolated based on the emission properties of the quasar to the red of the Lyα emission, using empirical models motivated by lower redshift observations.

To estimate the sensitivity of our abundance limits to systematic uncertainty in the continuum fit, we constructed five separate continuum models using different methods and assumptions and calculated the minimum and maximum $N_{HI}$ that could be consistent with the normalized data.

Three of these five models are scaled composite spectra of lower redshift quasars from the Sloan Digital Sky Survey. The first two are straight mean stacks of quasars in the SDSS which cover the Lyα line. The first[15] was derived from the full SDSS Data Release 3 sample, and the second[14] – which we use as our fiducial model – utilizes a somewhat larger sample with improved quasar redshifts[16]. We also examined a separate composite spectrum of SDSS quasars selected specifically to have large C IV blueshifts[14]. Given the very large velocity offset between Mg II and C IV observed in ULAS J1120 (Ref. 6) this spectrum (hereafter the "blue" continuum) is of particular interest.

The fourth continuum model is derived from a stack of 101 quasars observed with the Faint Object Spectrograph on the Hubble Space Telescope[17]. This continuum model suffers less from Lyα forest contamination than the SDSS examples on account of its lower mean redshift, but has broadly similar emission line profiles. We derived essentially identical results between this and our fiducial mean spectrum, with log ($N_{HI}$) differing by only 0.01 dex.

The fifth and final continuum model is determined empirically by fitting the quasar flux at $\lambda_{em} > 1235$Å to a basis set of quasar eigenspectrum templates derived through principal-component analysis[18]. We found that ULAS J1120's spectrum could be reproduced very well using just the first 5 basis vectors of the 50 dimensional set, indicating that its emission properties are quite typical of low redshift quasars. After the basis coefficients were determined, we extrapolated the reconstructed linear combination to include the Lyα emission region.

All of these continuum models are defined in the quasar's rest frame and must be redshifted to match the observed infrared spectrum. But while the quasar's emission redshift $z_{em}$ is known very accurately[25] from measurements of [C II] at z=7.0842, moving the template to this redshift will not necessarily provide the most accurate model of the Lyα line profile. This is because ULAS J1120's rest frame UV emission lines, most notably C IV and C III], are blueshifted by ~1700 km s$^{-1}$ relative to Mg II[6], and therefore Lyα may be as well. Emission line blueshifts are normal in quasar spectra[28], but the magnitude of the offset is unusually strong in ULAS J1120. Consequently, low-redshift composites made from objects with smaller emission line blueshifts do not reproduce the profiles of strong emission lines – including Lyα – for ULAS J1120. Instead, we adopted a separate template redshift $z_{templ}$ whose exact value varied between continuum models, but which was always at or slightly below $z_{em}$.

Among the different continuum realizations, the fiducial SDSS mean spectrum[14] provides the best match to the observed data for an adopted $z_{templ}$=7.07 (Figure 1). This combination produces an excellent correspondence to the data in the λ = 1.0-1.6 μm region. As expected, it fails to capture the blueshift of C IV and C III], but the equivalent widths and velocity spreads of all lines are correctly predicted. At this value of $z_{templ}$ the Lyα line is blueshifted by 527 km/s, much smaller than C IV for which Δν = 1670 km/s. Statistical studies of quasar emission line shifts have found that Lyα tends to be blueshifted, but less so than C IV by a factor of ~2 (ref 28). If this template were further blueshifted, its unabsorbed continuum estimate would fall below the observed flux in the N V region, which is clearly unphysical. Larger template redshifts are allowable; these universally lead to higher $N_{HI}$ measurements and thus lower inferred heavy element abundances.

The specially selected "blue" continuum described above provides the worst fit of all the models considered. This is partly by construction, because the quasars stacked into the blue template were jointly selected to have low emission line equivalent widths. Accordingly the template systematically underestimates the strengths of all broad emission lines in ULAS J1120 by a factor of ~2, including the Lyα/N V blend, where the continuum falls well below the data. Blueshifting the template exacerbates this problem, so we fix $z_{templ} = z_{em}$ for this model. Because the continuum falls below the data, we derive our lowest formal value of log ($N_{HI}$) = 20.35 for this model but consider it very unlikely on account of the poor quality.

The full SDSS composite[15] yields similar results to the fiducial spectrum[14] for template redshifts between z = 7.04 and 7.07. As with the fiducial spectrum, we find that $z_{templ}$ = 7.07 provides the best match to the full run between 1.0 and 1.6 μm. If we restrict to the Lyα/N V region and ignore a poor fit between 1.0 and 1.4 mm, a suitable solution may be found as low as $z_{templ}$=7.04, which achieves the minimum plausible log ($N_{HI}$) of 20.45 for this model.

Finally we consider the PCA template spectrum. Not surprisingly, this prescription produces acceptable fits for any template redshift between $z_{templ}$ = 7.035 and 7.085, since we are varying six free parameters (5 eigenmode coefficients plus redshift) rather than two (normalization plus redshift, as in the composite templates). The PCA basis spectra tend to produce larger Lyα emission line fluxes, resulting in high H I column density estimates. For all five continuum models, we investigated the effect of including a simple slope as an additional parameter to fit the spectral data. In some individual cases this led to a better fit or slight modification in $N_{HI}$, but the systematic error envelope quoted below on $N_{HI}$ includes all possible sloped or non-sloped fits.

In the analysis below, we adopt the fiducial mean quasar spectrum; among our choices it best represents the ULAS J1120 spectrum from 0.99-2.3 μm.

### S.3. H I Column Density Fitting

For each continuum model, we calculated the H I column density using the VPFIT software package to match Voigt profiles to the Lyα damping wing. We performed fits at fixed b = 10, 20, and 30 km s$^{-1}$ to verify that the specific choice did not affect our column density measurements. For our fiducial continuum, we derive log ($N_{HI}$) = 20.60 +/- 0.05, as shown in Figure S1. Although not shown in the figure, we performed joint fits to determine the redshifts and column densities of the two optically thin Ly α absorption lines at 0.981 μm and 0.9835 μm. Their arrangement resembles a doublet, but we ruled out the normal metal-line transitions seen in QSO spectra (C IV, Si IV, N V, Mg II, Al III, and permutations of Fe II) leaving Lyα as our favored identification. The best fit parameters for the system at 0.981 μm are $z_{abs}$ = 7.0721±0.0001 and log ($N_{HI}$) = 14.396 ± 0.095, corresponding to a velocity offset of -426 km/s (blueward) from the quasar's systemic redshift. The line at 0.9835μm has fit parameters of $z_{abs}$ = 7.0885±0.0001 and log($N_{HI}$)=14.161±0.033, corresponding to a velocity offset of +161 km/s (redward). Because the host redshift z = 7.0842 is well constrained from observed [CII] emission, the associated velocity offset of the reddest H I implies the presence of infalling or accreting gas onto the quasar host galaxy[40].

The redshift of the damped absorption was allowed to vary during our fits and converged to a value of $z_{abs}$ = 7.041 +/- 0.003. Since IGM absorption eliminates the blue wing of the profile, the line center is not strongly specified so we investigated the degree to which a change in column density could be effected by forcing a change in $z_{abs}$. By lowering $z_{abs}$ to 7.035 we increase the best fit column density to $\log(N_{HI})=21.0$, since more absorption is needed to match the damping wing. Likewise by increasing $z_{abs}$ we could lower $N_{HI}$, subject to the constraint that the transmission must be non-zero at 0.980 μm. At higher $z_{abs}$, $N_{HI}$ must be lowered to narrow the blackened line core and transmit at 0.98 μm. But to satisfy this constraint the damping profile must be substantially reduced, which then causes the fit to fail in the wings.

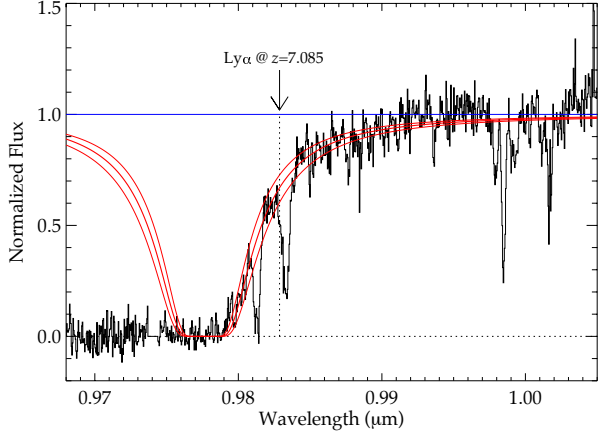

**Figure S1**: Continuum-normalized spectrum of ULAS J1120, with Voigt profile fit. The best fit model converges to $z = 7.041$ and $\log(N_{HI}) = 20.60$. The quasar spectrum has been normalized by the continuum model shown in Figure 1. The dotted vertical line indicates the location of Lyα at the quasar's rest frame. A N V absorption doublet at the quasar's emission redshift is visible near 1.00 μm. An instrumental artifact has been masked out at 0.974 μm.

Likewise, changes in $z_{templ}$ can introduce systematic changes in $N_{HI}$. This effect may be understood from Figure 1: at higher $z_{templ}$ the peak of the Lyα line moves into alignment with the damping wing region at λ=0.980-0.985 μm. The increased continuum-to-object contrast results in a more suppressed damping wing and correspondingly high column density. In the same way, lower values of $z_{templ}$ lead to systematically lower $N_{HI}$ values and weaker abundance limits.

A small region of pixels falling above the best-fit line is visible in Figure S1 near λ = 0.979 μm, which is clearly unphysical. However this region falls atop an OH sky emission line, so the difference between the model and fit are within the expected random errors. If we required the model to lie on or above these points, the best-fit redshift decreases and the H I column increases.

Over the full range of parameters tested, the best fits consistently fell in the range $\log(N_{HI}) =$ 20.55-20.65. Plausible combinations of $z_{templ}$ and $z_{abs}$ exist that yield extreme answers in the range 20.45-21.0, with the lowest $N_{HI}$ values coming from large $z_{abs}$ (limited by the non-zero flux at 0.98 μm) and small $z_{templ}$ (limited by the requirement that the continuum not fall below the object flux near Lyα + N V).

### S.4. Modeling the IGM near the Quasar's Ionized Near-Zone

The procedure described in Section S3 takes a purely empirical approach to fitting $N_{HI}$, which is appropriate if the damping wing is produced by a discrete DLA close to the QSO. Alternatively, the wing could result from extended Gunn-Peterson[30] absorption in a neutral IGM. In practice a luminous QSO in a partially neutral medium will create a zone of enhanced ionization, separated from the surroundings by an ionization front (IF, Ref 31). Residual H I inside the IF can still imprint a substantial absorption signature on the quasar spectrum[29]. Here we investigate a simple model of the ionized zone and surrounding IGM. Our goals in developing this model are: 1) to determine for what assumptions about the QSO lifetime and pre-existing IGM neutral fraction a damping wing will arise purely from the IGM, and 2) in the scenario of an ionized IGM+DLA, to quantify how much our DLA fit may be biased by not accounting for residual H I absorption in the near-zone.

We assume the QSO turns on in the midst of an IGM with pre-existing neutral fraction $f_p$, and has since remained in its active or bright phase for $t_Q$ years. The evolution of an ionization front in this scenario has been worked out in detail[31-33,29]; its radius evolves as

$$R_{ion} = \frac{4.2}{(\Delta f_p)^{1/3}} \left( \frac{\dot{N}}{2 \times 10^{57} \text{sec}^{-1}} \right)^{1/3} \left( \frac{t_Q}{10^7 \text{yr}} \right)^{1/3} \left( \frac{1+z}{7} \right)^{-1} \text{Mpc}$$

(S1)

where $\Delta$ is the local overdensity, and $\dot{N}$ is the (monochromatic) ionizing flux. Inside of $R_{ion}$ the gas is highly ionized and in photoionization equilibrium with radiation from the QSO; outside it retains the prior ionization condition of the IGM. Given the production rate $\dot{N} = 1.3 \times 10^{57}$ sec$^{-1}$ of ionizing photons from ULAS J1120 (Ref. 6), one may calculate the H I ionization balance as a function of distance from the QSO, whose flux falls as $R^{-2}$ due to geometric dilution. For the highly ionized zone behind the IF, the neutral fraction is

$$f_{nz} = \frac{4\pi \bar{n}_H \Delta \alpha_A}{\sigma_\alpha \dot{N}} R^2 \quad (S2)$$

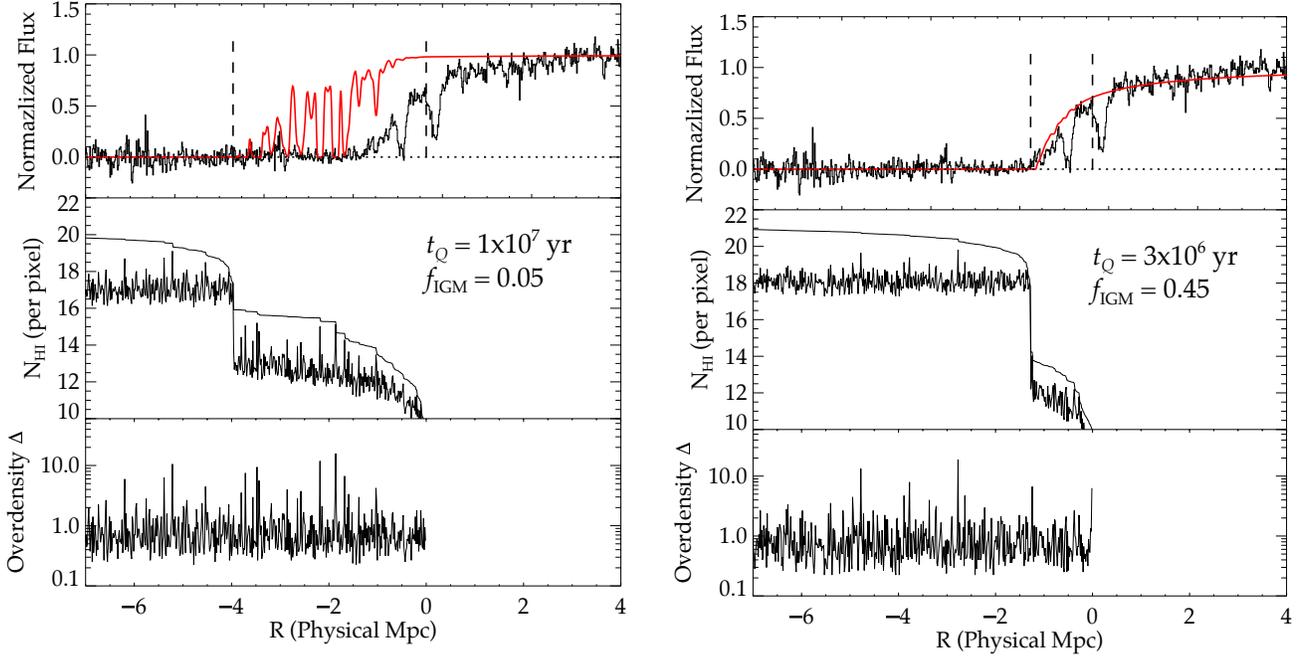

**Figure S2:** Near-zone models for an ionized (left) and substantially neutral (right) IGM. Red curves show typical Monte Carlo realizations generated using procedures described in the text, compared to the observed profile (black, top panel). The center panels indicate the H I column density vector in pixel units, and the cumulative column density from the QSO in the smooth curve. The bottom panels show overdensity vectors, drawn from empirical fits to simulations, and scaled to $z$ = 7. In the ionized model, additional absorption from a proximate DLA is required to reproduce the wing redward of the systemic Lyα wavelength (shown by the right dashed vertical line). Large neutral fractions can generate the damping wing from IGM absorption alone.

Where $\bar{n}_H$ represents the mean hydrogen density at $z$ = 7, $\alpha_A$ is the Case A recombination coefficient, and $\sigma_a$ is the H I photoionization cross section at 1 Rydberg. Given this ionization fraction and the local number density $\bar{n}_H \Delta$ of hydrogen, one may calculate the H I column density per pixel inside of the ionization front. The corresponding optical depth ($\tau = \bar{n}_H \Delta \sigma_\alpha f_{nz} dl$) is proportional to $\Delta^2$, and also $R^2$, implying that for a constant density field the absorption profile will have a Gaussian shape[29].

Because of the strong dependence on $\Delta$, we implemented an heuristic model of the density field to compare our simulated near-zone profiles with the data. Lacking access to numerical simulations, we used parametric fits to the probability density function (PDF) of $\Delta$ taken from simulations described in the literature[34]. After extrapolating and normalizing the PDF fit parameters for these models to the appropriate redshift, we generated Monte Carlo realizations of the density field from this weighting function. In utilizing a one-point statistic we will not capture correlations in the density reflecting power on varying scales, but we do represent the correct distribution of optical depths, since most volumetrically random samplings will have densities slightly below the mean.

Outside of the ionization front, the ionization state of the IGM remains unchanged, and is (at present) unknown. Following earlier precedent[29,5], we leave the volume-averaged $f_p$ as an input parameter. However we do allow the ionization fraction to vary with density. In this environment the usual approximation of $f_p \ll 1$ for the IGM does not necessarily apply, so one must solve the quadratic rate balance equation

$$n_H f_p \Gamma = (1 - f_p)^2 n_H^2 \alpha_A \quad (S3)$$

where the ionizing photon rate Γ is driven by the production of 1 Ryd photons from early galaxies of unknown number density, spatial distribution, and escape fraction. It is beyond the scope of this paper to address the nature of these sources; instead we use the average value of $f_p$ (an input to the model) and $\bar{n}_H = \Omega_b \rho_c (1+z)^3 / \mu m_H$ to calculate an average value of Γ. Then, we fix Γ at this constant value and solve for $f_p$ as a function of the local overdensity Δ.

Our model deliberately excludes H I self-shielding from the radiation field, which suppresses the production of damped Lyα absorbers from high density regions[5]. We made this choice to artificially force the damping wing to derive entirely from the diffuse IGM. If DLAs are injected back into the near zone by including self-shielding the abundance calculation would revert back to the discussion of Sections S3 and S5.

With this recipe we derive a vector of $N_{HI}$ per spectral pixel spanning the ionization front, for each Monte Carlo realization. To generate an observed absorption profile, we convolve this vector with a Voigt function of width $b$ = 20 km s$^{-1}$ and then further convolve with FIRE's instrumental profile.

An example profile for $f_p$ = 0.05 and $t_Q$ = 1 x 10$^7$ years is shown in the left panel of Figure S2, along with the corresponding vectors of overdensity Δ and $N_{HI}$ per pixel. Clearly this model – which does not include a DLA – fails to reproduce the large optical depth seen near the quasar's redshift. In particular, it does not generate absorption redward of the systemic Lyα wavelength, marked with a vertical dashed line at right. The right panel shows that for a short QSO bright phase and high IGM neutral fraction, it is possible to reproduce the damping wing profile with no additional absorption, consistent with prior studies[5,6]. Notably, in this model the cumulative $N_{HI}$ within ~2 Mpc of the IF approaches 20.6 in the log, the value determined for our single-component DLA fit.

We will address metal absorption signatures of the IGM-only profile below; here we consider whether patchy H I in the near-zone could bias our $N_{HI}$ measurements for the ionized IGM + DLA case[35,36,29]. Given a single statistical realization of the near-zone, it is straightforward to calculate the amount of additional DLA absorption required to match the observed spectrum by treating the red curve in Figure S2 as an effective continuum for a Voigt profile fit. By iterating this procedure over many Monte Carlo iterations with varying input parameters (300 per combination of $f_p$ and $t_Q$) we developed distributions of $N_{HI}$ for the DLA, shown in Figure S3. We conservatively chose not to mask out narrow, strong features in the continuum as would be done on a real fit, although we did include two additional H I components for the known narrow lines in the observed near-zone.

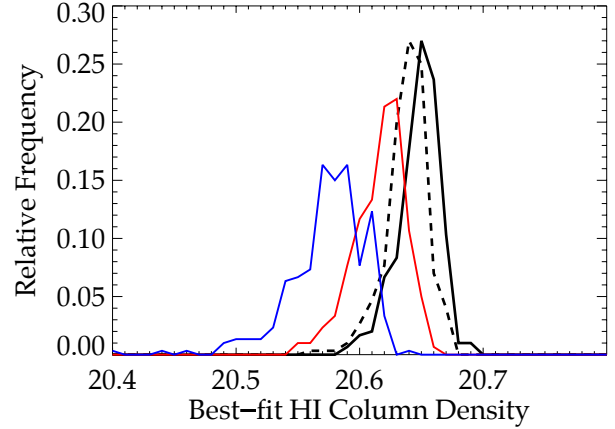

**Figure S3:** Distribution of DLA best-fit column densities when residual, patchy absorption inside the quasar's near-zone is considered. The solid and dashed black curves are for $f_p$=0.01 and $t_Q$=1 x 10$^7$ and 1 x 10$^6$ years, respectively. The colored curves both have $t_Q$ = 1 x 10$^7$ years, but neutral fractions of $f_p$=0.05 for red and 0.10 for blue. With increasing neutral fraction and decreasing QSO lifetime, the effective column density decreases, but only at the level of 0.02-0.10 dex.

The black histograms show that for an IGM neutral fraction of 1% and a range of quasar lifetimes, the best-fit $N_{HI}$ including near-zone absorption differs from the simple case by only 0.01 - 0.02 dex in the mean, with a scatter of σ ~ 0.02 dex. As $f_p$ increases to 5% and 10% (red and blue curves), the bias grows to 0.03 dex and 0.07 dex, but the likelihood of log($N_{HI}$) < 20.5 remains relatively low. For combinations of $f_p$ > 10% and $t_Q$ ~ 10$^6$ years, the meaning of a distinct DLA component becomes ambiguous, because the profile approaches the Damped IGM state shown at right in Figure S2. But for near-zone models with a DLA embedded in an ionized IGM, the effect of residual H I on abundance measurements appears to be small.

*S.5. Heavy Element Abundance for a Single-Component DLA*

Given that the Lyα profile may be fit either as a discrete DLA in a largely ionized IGM, or as the summed contribution of a continuous and substantially neutral IGM, we must consider the signatures of metal absorption in two separate cases. We discuss the discrete DLA case here and address the IGM case in the following section. For the DLA model, we adopt the best-fit parameters for the H I profile of $z = 7.041$ and $\log(N_{HI}) = 20.6$.

For these parameters, we estimated upper limits for the column density of several key heavy element transitions using both a curve-of-growth analysis and Voigt profile modeling. We calculated rest-frame equivalent width limits for each expected line using an unweighted sum of the normalized spectral pixels, over an aperture spanning $+/-2\sigma$ from the line center. The width $\sigma = 21$ km s$^{-1}$ was derived from the instrumental resolution profile of FWHM 50 km/s. For all transitions the rest frame equivalent width ($W_r$) measured in this way is less than its $1\sigma$ error (except for O I, which shows absorption at the $2.2\sigma$ level). We report $1\sigma$ and $2\sigma$ upper limits for each ion, where quoted $n\sigma$ limit is calculated as max[$n\sigma$, $W_{r,\ measured} + n\sigma$].

These limits, which range between 10 and 100 mÅ, are listed in Table S1. In most cases they restrict the location of any underlying absorption to the linear, unsaturated portion of the curve-of-growth except for small choices of the Voigt $b$ parameter. Curves of growth for the various transitions are shown in Figure S4, where we have indicated with vertical lines the $1\sigma$ and $2\sigma$ limits on $W_r$ and with horizontal lines the derived column densities for $b = 10$ km/s. For our $1\sigma$ limits, saturation effects are negligible for $b > 10$ km/s, and only amount to 0.1 dex at $b = 5$ km/s. For our $2\sigma$ limits the effect remains negligible until $b$ approaches 5 km/s, but can introduce errors of 0.2 dex in some abundances. Si II, which provides our most stringent metallicity limit, is unaffected by saturation. The column density limits listed in Table S1 are derived using $b = 10$ km/s but may be scaled as appropriate from Figure S4 to account for saturation at lower $b$.

To check the curve of growth calculations, we also fit upper limits to the heavy element column densities directly using the VPFIT software package. For each transition we fit pixels within the same aperture described above, centered on $z = 7.041$. Lacking an actual measurement of $b$ we again fixed its value at 10 km s$^{-1}$. The fitted column density limits generally agree with those derived from curve-of-growth at the 0.1 dex level. With these column density estimates, we derive abundances for each element as [X/H] = $\log(N_x) - \log(N_{HI}) - \log(X/H)_\odot$; using $\log(N_{HI}) = 20.6$, and normalizing our abundances to a standard solar scale.[24]

Although most stellar and DLA abundance studies use [Fe/H] as a classification benchmark, our most sensitive limits are derived from [Si/H] on account of its clean spectral region and large oscillator strength. The [Fe/H] ratio is most tightly constrained by the 2600Å line, from which we derive [Fe/H] < -3.41 ($2\sigma$).

| Ion | $W_r$ [Å] | N (COG) [cm$^{-2}$] | N(VPFIT) [cm$^{-2}$] |
|---|---|---|---|
| O I 1302 Å | ≤0.044 (0.057) | ≤13.90 (14.04) | ≤14.04 (14.16) |
| C II 1334 Å | ≤0.017 (0.034) | ≤13.00 (13.34) | ≤12.84 (13.14) |
| Si II 1260 Å | ≤0.009 (0.018) | ≤11.76 (12.08) | ≤11.86 (12.16) |
| Fe II 2586 Å | ≤0.038 (0.054) | ≤13.01 (13.18) | ≤13.12 (13.31) |
| C IV 1548 Å | ≤0.005 (0.009) | ≤12.07 (12.38) | ≤12.57 (12.87) |
| Mg II 2796 Å | ≤0.072 (0.096) | ≤12.31 (12.46) | ≤12.37 (12.52) |
| Fe II 2600 Å | ≤0.059 (0.090) | ≤12.68 (12.90) | ≤12.67 (12.69) |
| "Stack" | ≤0.014 (0.020) | ≤13.32 (13.49) | ≤15.48 (15.78) |

**Table S1:** Heavy element absorption upper limits for a discrete DLA at z = 7.041. For each ion, we list the $1\sigma$ ($2\sigma$) limits on the rest-frame equivalent width, and its associated column density derived using the curve-of-growth (COG). Finally we show a direct limit on each column density obtained using VPFIT. All limits were measured over an aperture of total width 84 km s$^{-1}$ centered at z = 7.041.

Referring to the literature on metal-poor Galactic halo stars[37], this classifies the DLA in ULAS J1120 as a candidate gas reservoir for "Extremely Metal Poor" stars, often defined as [Fe/H] < -3. However the limits on [Si/H] are over three times lower at [Si/H] < -4.03 ($2\sigma$). While we cannot measure the relative abundance between elements with upper limits alone, if we assume a value of [Si/Fe]=0.3, typical of metal poor DLAs from the literature[12], the implied [Fe/H] would approach -4.3, in the range of the "Ultra Metal Poor" stars. The upper limits to [C/H] and [Mg/H] are likewise in the -3.75 to -4.0 range.

We do see evidence of heavy elements in a highly ionized phase near ULAS J1120, but these lines are at substantial velocity separation from the damped Lyα absorption and are not likely to be physically related. As reported previously[6], there is intrinsic C IV and N V absorption offset from the $z = 7.041$ neutral H I by $\Delta v$ = +600 to +800 km s$^{-1}$; we also see weak evidence of C IV absorption at $z = 7.016$ ($\Delta v$ = -1000 km/s from the H I absorption; our error in the DLA redshift is $\Delta v \approx \pm 100$ km s$^{-1}$). These systems do not show any low-ionization lines (e.g. C II, Si II, O I, Fe II), and for the higher redshift system the H I absorption is unsaturated in Lyα. The two metal absorbers and the DLA all fall within 3000 km s$^{-1}$ of the background quasar, an area normally excluded from absorption line surveys on account of local ionization and enrichment, and interloping absorption from outflowing material near the quasar. If one wishes to attribute velocity differences to Hubble flow, it may be possible to construct a scenario involving patchy enrichment on ~1 Mpc scales to explain these metals, but in this case one must also explain their high ionization state. In particular, if the candidate system at z = 7.016 is indeed closest to earth, then the ionizing flux from ULAS J1120 would be significantly attenuated by the intervening DLA and a second, closer source of hard photons may be required. In contrast the highly ionized metals are explained very naturally as material close to and ionized by the quasar, but flowing earthward at 1000-3000 km s$^{-1}$.

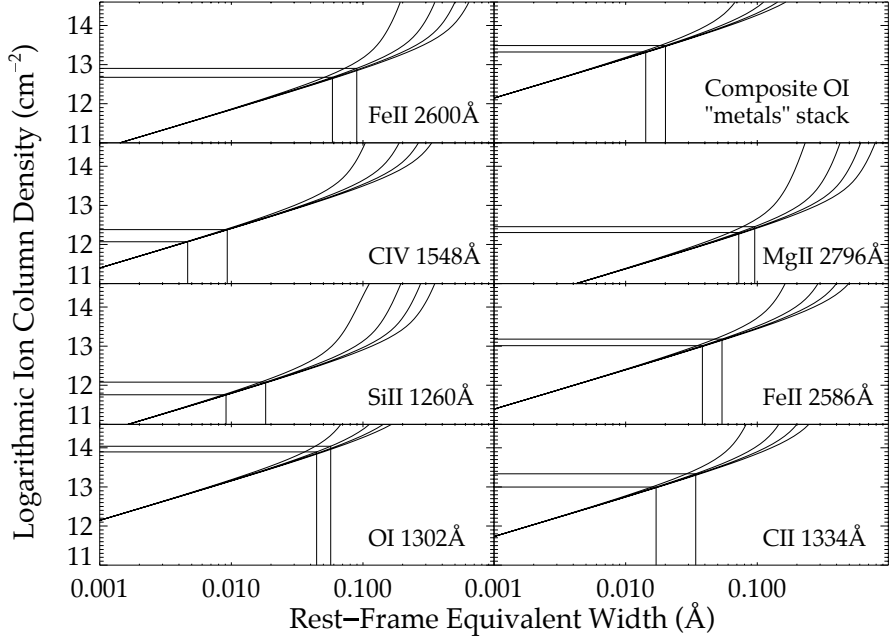

**Figure S4**: Curve of growth analysis to derive column densities for selected ions. Limits from the stacked metal line composite are also shown at upper right. For each panel, 1σ and 2σ limits to the equivalent width and column density are shown with vertical and horizontal lines, respectively. The curves represent *b*-parameters of 5, 10, 15, 20 km/s from left to right in each panel.

*S.6. Heavy-Element Abundance in the Case of Extended IGM Absorption*

The low chemical abundances derived above require that the absorption arises in a single, gravitationally collapsed structure. If the damping wing is instead caused by Gunn-Peterson absorption from the IGM, the absorbing gas would be spread over Mpc scales and its associated heavy element signature would be similarly extended. In this section, we develop our model of the QSO near zone to incorporate heavy elements, and consider what abundance limits are consistent with the data in this configuration.

We generated Monte Carlo near-zone profiles using the methods described in Section S4, covering a range of metallicity from [X/H] = -4.0 to [X/H] = -2.5 in steps of 0.25 dex. For each realization, heavy elements were added with a spatially constant value of [X/H] and solar relative abundances. As with H I, fluctuations in the overdensity Δ lead to corresponding line-of-sight variations in the heavy element optical depth – a metal-line "forest."[38] For consistency with the observed shape of the Ly α + IGM damping profile in Figure S2, we used $f_p$ = 0.45 and $t_Q$ = 3 x 10$^6$ years.

The ionization balance of heavy element lines outside the ionization front differs slightly from hydrogen. These regions do not yet "see" hard ionizing photons from ULAS J1120, so at z = 7 the metagalactic background radiation spectrum should be dominated by galaxies. This source spectrum is inherently soft, and would be strongly attenuated above the Lyman limit by ambient hydrogen since the IGM is (by construction) not reionized. The metal line transitions considered here (Mg II, C II, Si II, Fe II) all require photons above the Lyman limit (1.1, 1.8, 1.2, 1.2 Ryd, respectively) to reach the doubly ionized state. With the volume-averaged H I fraction near 50%, the mean free path of such photons should be very small, so these elements will all remain singly ionized. We confirmed this using CLOUDY simulations with a Haardt & Madau ionizing spectrum of galaxies plus QSOs at $z = 7$; these suggest that ionization corrections would only be of order 0.1 dex (though this should only be viewed as a proof-of-concept since a spatially uniform background oversimplifies the radiation field during reionization). The one exception is O I, which differs in ionization potential from H I by only one part in $10^4$. In the discussion below we tie the ionized fraction of O I to that of H I, and assume that Mg, C, Si, and Fe are entirely in the singly ionized state outside the ionization front. Inside the ionization front, these transitions are ionized away and generally not visible.

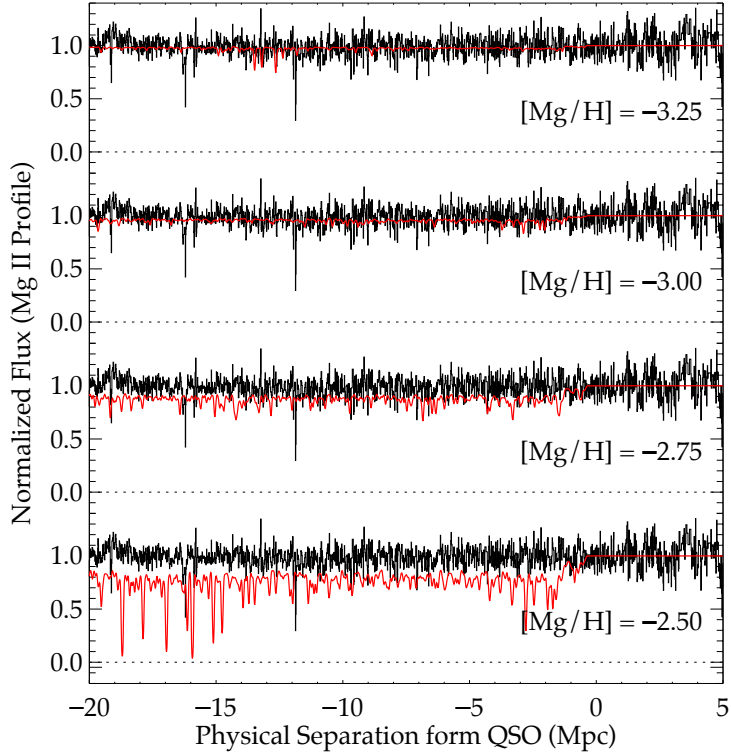

**Figure S5:** Monte Carlo realizations of the Mg II absorption profile for an IGM with 45% neutral fraction (consistent with the HI profile) and varying degrees of heavy element enrichment. Red curves illustrate the simulated normalized flux profiles, black curves show the normalized spectrum of ULAS J1120. Absorption is prominent for [Mg/H] > -2.75, but statistical methods are needed to infer signal at 0.001 Solar and below, and continuum fitting errors must be taken into account.

Combining our estimates of the mean hydrogen density, statistics of the overdensity[35], and the above assumptions about ionization, we generated vectors of heavy element column density for each IGM pixel. As before, we convolve the column density vector for each metal line with a Voigt profile of 20 km s$^{-1}$ and then the FIRE line spread function to obtain the optical depth vector for each Monte Carlo trial. Figure S5 shows examples of four such realizations for Mg II with varying metallicity, superimposed on the observed spectrum. Simple inspection suggests that with an accurate continuum estimate, absorption at the [X/H] > -3.0 level should be readily detectable, while at [X/H] < -3.0 the metal line signal blends into the noise.

This may be understood qualitatively by considering the optical depth at line center for an isolated, single Voigt profile

$$\tau_0 = \bar{n}_H \Delta \left(\frac{X}{H}\right)_\odot 10^{[X/H]} \left(\frac{\pi e^2}{m_e c} f_{12} \frac{\lambda_0}{\sqrt{\pi} b}\right) dl \quad \text{(S4)}$$

which evaluates to $\tau_0 = 0.02\Delta$ for Mg II at $z = 7$, [Mg/H] = -3.0, and $dl$ = 14 kpc pixel$^{-1}$ for FIRE. Such a 2% decrement represents a challenging detection for data with SNR ~ 10, and may only be established statistically. For regions with $\Delta$ > 5-10, or higher chemical abundance, a >10% decrement is possible, enabling the detection of discrete lines. However our abundance limits for diffuse IGM absorption will not be as stringent as in the DLA configuration; detection of localized absorption with [Mg/H] = -4.0 (a 0.2% effect at the mean cosmic density) is presently beyond reach.

Figure S5 suggests that statistical searches for heavy element absorption from a neutral IGM could trigger on two possible signatures. In the idealized case where the metals are uniformly distributed, there is a discrete step discontinuity in optical depth at the edge of the ionization front, and one can leverage many pixels to increase the edge detection sensitivity. However it seems likely that metals would not be distributed uniformly at such early times, and would instead be concentrated into overdense regions. In this scenario it is more appropriate to search for discrete lines using traditional methods.

We first tested the step-detection method by generating 1,750 near-zone metal line profiles for the heavy element ions. We multiplied these absorption profiles into the actual ULAS J1120 data and stacked the resulting near-zones for each ion into a composite metal line profile. Then, we calculated the mean optical depth at $1 < R < 6$ Mpc blueshift from the QSO. We excluded the 0-1 Mpc region because it is inside the ionization front for choices of $f_p$ and $t_Q$ that match the Lyα profile. A reference optical depth was established using a similar-sized region redward of the QSO emission redshift, and the reported value is the ratio of the test zone to this reference.

Figure S6 shows the derived mean optical depth as a function of metallicity. For comparison, the mean optical depth of ULAS J1120 (without addition of simulated absorption) is shown with a thick dashed line. The blue curve, which assumes a perfect continuum fit, indicates that absorption at the easily detectable 5-15% level is present for metal abundances above [X/H] > -3.25. Below this level there is additional signal, but systematic errors at the 1-5% become a source of concern, especially regarding the continuum fit.

In data of SNR ~ 10-20, a dip in the profile from diffuse metal absorption could easily be fitted over when estimating the QSO continuum, diluting true absorption signal. To test the effect of this potential bias, we re-ran the mean optical depth analysis on all 1,750 sightlines, but with continuum fits recalculated from scratch for each sightline after adding the simulated absorption. We performed the fits using the PCA continuum code[18] described in Section S2 with 5 basis functions and iterative masking of outlying pixels. Importantly, this continuum is determined globally by fitting orthogonal QSO basis templates, rather than by local fit of splines. By design, this is less susceptible to over-fitting high spatial frequency

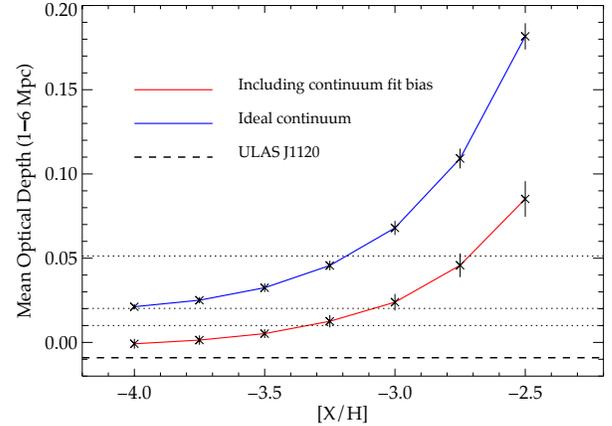

**Figure S6**: Measured step discontinuity from the blue to red side of the QSO emission redshift for the stacked metal absorption profile, at different abundance levels. Blue curve assumes perfect knowledge of the continuum, and the red curve incorporates the effect of continuum fitting bias induced by additional heavy element absorption. Heavy dashed line indicates the measured value for ULAS J1120, while the light dotted lines indicate errors at the 1, 2, and 5% level. The ULAS J1120 spectrum has SNR ~ 10-20 per pixel, but ~800 pixels are averaged to measure the optical depth.

features in the continuum, though no procedure is fully immune.

The results with biased continuum fits are shown in the red curve. Even with the global PCA fits, the mean optical depth is systematically underestimated by a factor of ~2, with a larger bias at large abundance, as expected. The difference suggests that metal absorption could affect continuum estimates at the ~5% level. However, a >2% discontinuity is still present at the [X/H] ~ -3.0 to -3.25 level. At [X/H] < -3.5 heavy elements in the diffuse IGM would remain largely undetectable via step discontinuity.

We also performed conventional searches for individual metal lines, as would be appropriate if the spatial distribution of metals was non-uniform. Our search algorithm convolves the absorption-injected data and error arrays with a Gaussian kernel of width one spectral resolution element. We then search the ratio of these arrays for peaks at signal-to-noise > 3 per resolution element. For sufficiently high values of Δ, metal lines are visible even for low heavy element abundances (See Figure S5). We wish to quantify the probability of encountering one of these detectable peaks within 5 Mpc of the ionization front, where the IGM damping wing would be generated. Figure S7 shows these probabilities drawn from our suite of Monte Carlo runs, for both the unbiased and biased continuum.

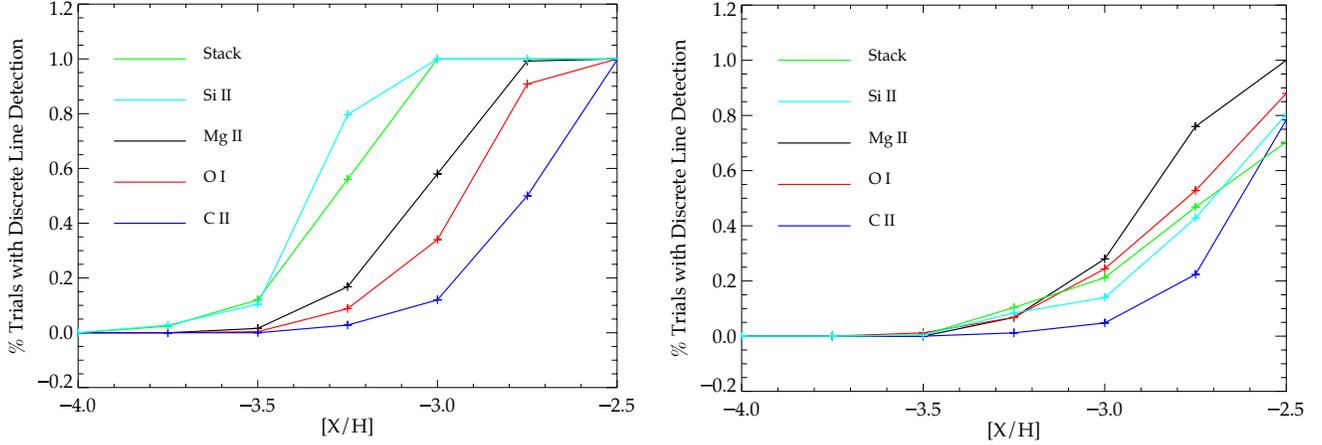

**Figure S7:** Estimated probability of detecting discrete metal absorption lines from density fluctuations in a neutral IGM. This is calculated as the fraction of near-zone realizations for which at least one discrete metal line is detected at significance >3σ per resolution element, within 5 Mpc of the ionization front. Left panel is for a best-case continuum fit, and right panel is the worst-case continuum bias. The true sensitivity depends on the spatial distribution of heavy elements; for sparse/patchy distributions the fit should resemble the idealized case, while for uniform metallicity it should approach the right panel, where step discontinuity detection becomes more sensitive.

For the unbiased continuum, the odds of a line detection are >75% at [X/H] = -2.75, and drop to 50% between [X/H] = -3.0 to -3.5, with sensitivity varying between ions. When continuum fitting bias is included as above, the detection probability declines, by an amount that varies with ion, but can be large (shown at right). In this most pessimistic scenario, we expect a high probability of line detection for [X/H] = -2.5 but a rapidly declining success rate to near zero at [X/H] = -3.5.

Our search window encompassed a narrow redshift range near the QSO; this was enforced to ensure that the metals probe the same volume that produces the observed H I damping wing. If the IGM remains neutral at larger separations then the probability of metal line detection increases in proportion to the pathlength increase. However at lower redshift the IGM may or may not remain neutral – it contributes little to the damping wing and could therefore be substantially ionized and still saturate in Lyα[36,37,29]. Also, if the enrichment of the IGM is substantially patchy or sparse, then the biased version of our continuum fit (which incorporates uniform metal absorption throughout the IGM) may be too pessimistic. The true value would then lie between these extremes and possibly closer to the ideal case.

Cosmological simulations with chemodynamics would be needed to resolve this question fully.

Taken together, these tests show that our sensitivity to diffuse metal absorption from the IGM depends on assumptions about how the heavy elements are distributed, how they are measured, and how much bias is folded into the continuum estimate. For a wide range of assumptions, we are able to detect heavy elements at or above [X/H] ~ -3.0, but are less sensitive at lower abundances. For spatially uniform heavy element distributions the continuum bias becomes more severe, but the step-discontinuity method becomes more effective and can push below $10^{-3}$ solar. For more patchy metallicity fields, the continuum bias becomes less severe and line searches become more efficient.

### S.7. Ionization and Depletion Considerations

In some DLAs it is possible to measure incorrect abundances if either partial ionization of the neutral species or depletion of metals onto dust grains is not considered. For comparably low values of $N_{HI}$, prior studies suggest that dust depletion should not be a significant source of error for this system[39] particularly given its low abundance in non-refractory elements such as oxygen.

However, given the absorbers' proximity to the most luminous known object in the universe at $z > 7$, we must consider the potential effect of local ionization on its derived abundances. The few studies of proximate DLAs at lower redshift find mixed results on the importance of ionization. Studies of the Al III/Al II ratio for proximate systems do not show a systematic distinction between proximate DLAs and their field counterparts, although the Si IV/Si II ratio may be slightly elevated in the same systems[12].

However ionization considerations are particularly germane if the H I damping wing has an IGM origin. Our calculations in Section S4 support the earlier work of Bolton et al[5], who concluded that the observed Lyα profile could be consistent with near-zone IGM absorption if it has a volume-averaged H I fraction of >10% and the QSO's active phase has been short. If up to 90% of hydrogen is ionized it is possible that our targeted heavy element transitions would be as well.

To test these effects, we ran a simple CLOUDY simulation and estimated abundance ionization corrections as a function of the hydrogen neutral fraction. The simulation used a plane-parallel gas slab with Solar relative abundances, illuminated by radiation from a nearby AGN-like continuum with power law slope $\nu^{-1.5}$ in the UV region, similar to existing studies of the same object.[5] In this framework we adjusted the ionization parameter to generate clouds with total $\log(N_{HI}) = 20.6$ and a range of neutral fractions. Figure S8 (top) shows abundance corrections derived from the output ionization balances of the elements studied here. Evidently the corrections are very small (<0.05 dex) even at neutral fractions 1-10%, with the exception of Si II, which has a correction of 0.1-0.15 dex, and Mg II, which can be affected at the 0.2-0.3 dex level. With the exception of Mg II, all of the corrections would push our abundance limits further downwards. This is to be expected because these transitions have ionization potential very close to, but slightly larger than 1 Rydberg, so their ionization will track H closely but lag slightly for any ionizing spectrum that declines toward higher frequencies.

If we allow the gas to have a non-negligible ionized fraction, then we can place extremely strong upper limits on the abundance based on the non-detection of C IV in our data. The bottom panel of Figure S8 shows the C IV column density predicted by our CLOUDY model for varying [C/H] and volume-averaged H I fraction. The shaded region is excluded by our measured upper limit on C IV. For neutral fractions of 10% or lower and $\log(N_{HI})=20.6$, [C/H] must fall below the -5.0 to -6.0 range to avoid a detection.

We conclude that ionization effects, if present, would only affect abundance estimates at a level below other uncertainties in the data. If higher signal-to-noise ratio observations reveal heavy element lines then this question may be revisited, but our non-detection of high ionization lines rules out the possibility of a large, unaccounted metal mass in a highly ionized phase.

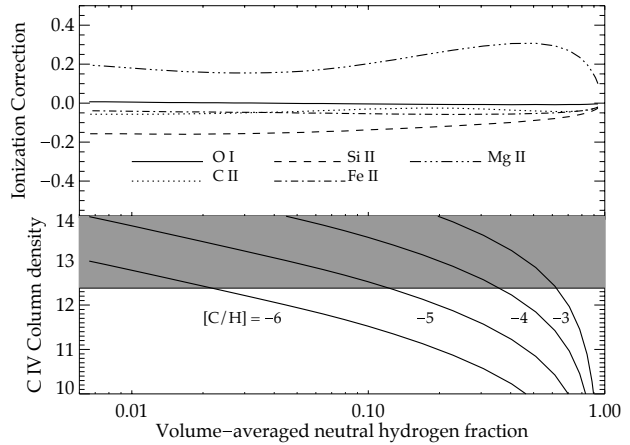

**Figure S8**: Exploration of the possible effects of absorber ionization. The top panel illustrates ionization corrections (as an additive adjustment in dex) to the abundances in Table 1 if the absorption arises from a partially ionized medium. The bottom panel shows limits on [C/H] for the ionized case. The shaded region is excluded by the observational upper limits on C IV absorption.

*S.8. Comparison to Prior Studies of Metal-Poor Quasar Absorbers*

Taken together, the arguments developed above indicate that our ULAS J1120 spectrum constrains the heavy element abundance at z = 7.04 to roughly Z < $10^{-3}$ Z$_\odot$ if the absorption occurs in diffuse intergalactic gas, or Z < $10^{-4}$ Z$_\odot$ if it occurs in a gravitationally collapsed DLA embedded in an otherwise ionized IGM.

At lower redshift, the least systematically enriched environment is the intergalactic medium. Previous studies using line fitting for highly ionized C IV and O VI[22], and pixel-optical-depth studies[23], have measured intergalactic abundances down to [C/H] ~ -3.4 with median -3.1, and an [O/H] that is 0.2-0.4 dex higher at z ~ 2.5. At z ~ 4 - 4.5, C IV is detected in individual Lyα forest absorbers with [C/H] ~ -3.4; a lognormal fit to the distribution implies a median abundance of [C/H] = -3.55[22]. The lack of heavy element signature in ULAS J1120 when the universe was 750 Myr younger is consistent with an IGM origin of similar or lower median abundance in the early universe. In this interpretation, the absence of metals is taken as confirmation that the absorber is intergalactic, implying that by z ~ 7 our ULAS J1120 observations are reaching well into the epoch of reionization. If so, the pattern of damped H I and low heavy element abundance should be generic to the spectra of z > 7 QSOs.

If the damping wing in ULAS J1120 reflects a collapsed structure, it would be the most metal-poor DLA yet known, and rank among the most chemically pristine environments in the known universe. The literature contains two examples of comparably low gas phase metallicities[4], both drawn from a sample of Lyman limit systems at z ~ 3 (1.4 Gyr after z = 7). The lower $N_{HI}$ associated with these absorbers results in a weaker absolute limit to the abundance, but also justifies a significant ionization correction. After this model correction, one obtains an upper limit of $Z < 10^{-4}$ $Z_{\odot}$, similar to what we quote. However these two examples are atypical of Lyman limit systems at z = 3. A compilation of the ~20 known Lyman limit systems with reliable abundance estimates yields an average of $10^{-1.4}$ $Z_{\odot}$, with full maximum and minimum ranging from $10^{-2.8}$ $Z_{\odot}$ to super-solar. Hundreds more Lyman limit systems have been observed but not corrected for ionization; these nearly all show metal line transitions, highlighting the unusual nature of the metal-poor LLS at z = 3. In contrast, the absorption system presented here is the first astrophysical environment (outside of an AGN broad line region) with a detailed abundance characterization at z > 5, and its metallicity is already among the lowest measured at any epoch.

### S.9. The Population III Abundance Threshold

If the absorption arises in a bound halo, then the age and low metallicity of this system suggest a possible connection to the old, metal-poor stars discovered in the halo of the Milky Way[1,2]. These objects have iron abundances below [Fe/H] = -5 but often exhibit extremely high [C/Fe] ratios approaching +4. With such a high carbon abundance a gas cloud of this composition would have been detected as C II in our absorber. However Frebel et al argue that the high [C/Fe] may reflect a selection effect particular to halo star samples, since only the lowest mass subset of stars formed at early times will survive to the present[8]. Low mass cloud fragmentation requires the existence of some cooling channel which is naturally provided by carbon fine structure lines in gas with elevated [C/H]. By this argument stars forming in the early universe with "normal" [C/Fe] ratios would not survive to be observed today.

By directly observing gas in the early universe, we remove this cooling/survivability requirement, and can potentially access reservoirs for low-metallicity star formation that lead to a Population III-like IMF. The boundary between the Population III and Population II modes has been summarized in terms of a single index $D_{trans} = \log_{10}(10^{[C/H]} + 0.3 \times 10^{[O/H]})$ (Ref 8). This quantity is calibrated to simulations of star formation at low abundance in the early universe. These simulations find that systems with $D_{trans} < -3.5$ tend to produce Population III stars; above this threshold cooling is efficient enough to support the Population II mode. We derive upper limits of $D_{trans} < -3.71$ (-3.51) at the 1σ (2σ) level for the z=7.041 absorber in ULAS J1120 (in the DLA configuration). This lies at or marginally below the metallicity threshold where Population III star formation is thought to take place.

Accurate theoretical models of low-metallicity cooling and fragmentation are still being refined. Indeed there appear to be low mass, metal-poor stars in the galactic halo[3] with no carbon enhancement; this finding requires either explanation in the context of existing models, a secondary cooling agent such as dust, or possible revisions to the fine-structure cooling models themselves. Nevertheless, it seems that near abundances of $10^{-4}$ solar, one experiences qualitative differences in the observed properties of stars, whether it be abnormally small lithium abundances and large variations in carbon enhancement (as have been observed) or a change in the shape of the IMF (as has been predicted).

The nearby quasar host, which exhibits heavy elements in both its broad emission lines and narrow C IV and N V absorption, demonstrates that localized chemical enrichment has begun by z = 7, at least in the most highly biased environments in the universe. However, even in the nearby foreground absorption system studied here, it appears that heavy element pollution has either not begun or has only proceeded to a level where stochastic variance in abundances could affect the properties of any stars formed. Large variations in the metal distribution on small scales are expected during the epoch of reionization. If crucial further abundance measurements continue to find low or variable metallicity it would suggest that we have begun to access the epoch of early stellar nucleosynthesis.